\documentclass{article}
\usepackage[utf8]{inputenc}
\usepackage{amsmath}
\usepackage{amsfonts}
\usepackage{braket}
\usepackage{graphicx}
\usepackage{physics}
\usepackage{xcolor}
\usepackage{hyperref}
\usepackage{multirow}
\usepackage{adjustbox}
\usepackage[a4paper, total={6in, 8in}]{geometry}
\usepackage[labelformat=empty]{caption}
\title{Momentum domain polarization probing of forward and inverse spin Hall effect of leaky modes in plasmonic crystals}
\author{Jeeban Kumar Nayak$^{1*}$, Harley Suchiang$^{1}$, Subir Kumar Ray$^{2}$,Ayan Banerjee$^{1}$,\\ Subhasish Dutta Gupta$^{1,3,4}$ Nirmalya Ghosh$^{1}$\\
$^{1}$\textit{\small Department of Physical Sciences},\\ \textit{\small Indian Institute of Science Education and Research Kolkata,}\\ \textit{\small Mohanpur, India- 741246}\\
$^{2}$\textit{\small Elmore Family School of Electrical and Computer Engineering, Purdue University}\\
$^{3}$\textit{\small Tata Institute of Fundamental Research Hyderabad, India}\\
$^{4}$\textit{\small School of Physics, Hyderabad Central University, India}\\
$^{*}$\textit{\small jkn19rs027@iiserkol.ac.in}}
\date{}
\begin{document}

\maketitle
\section*{Abstract} 
Simultaneous manifestation of both forward and inverse photonic Spin Hall effect in geometrically tailored anisotropic waveguided plasmonic crystal system is observed through the excitation of the leaky hybridized quasiguided modes. The quasiguided modes of the plasmonic crystal manifested in the far-field momentum (Fourier) domain as circular ring-like intensity distributions, and the different spin orbit interaction (SOI) effects revealed their exclusive signature as polarization-dependent azimuthal intensity lobes on top of it. Using a darkfield Fourier domain polarization Mueller matrix platform, we have observed input spin (circular polarization)- dependent trajectory of the leaky quasiguided modes in the momentum domain (forward spin Hall effect) and its reciprocal effect as the wave vector-controlled spin selection of quasiguided modes (inverse spin Hall effect). These effects are separately manifested in characteristic Mueller matrix elements enabling their interpretation as geometrical circular anisotropy effects. Resonance-enabled enhancement and control of these effects are also demonstrated by exploiting the spectral Fano-type resonance. The far-field manifestation of spin-directional excitation of leaky quasiguided modes, their unique interpretation through momentum domain Mueller matrix, regulation and control of the SOI effects in plasmonic-photonic crystal systems, opens up exciting avenues in spin-orbit photonic research.\\
\\
\\
\section*{Introduction}
Coupling and mutual influence of the spin, orbital angular momentum (AM), and linear momentum degrees of freedom of light have led to a number of fundamental photonic SOI effects in various light-matter interactions \cite{nphotonbliokh2015spin,PSOIwielly}. Spin and orbital Hall effect of light \cite{photonicshe,bliokh2015quantum,ohe}, optical Rashba effect \cite{rashbascience}, plasmonic Aharonov–Bohm effect \cite{aharonov-bohm}, spin-dependent transverse momentum, transverse SAM\cite{transverse}, spin-momentum locking and spin controlled directional waveguiding etc.\cite{soiwaveguide,soiwaveguide2} have provided exciting opportunities to gain new insights on the universal SOI phenomena in relatively simple classical optical settings. These have also opened up an entirely new paradigm of spin-orbit photonic devices\cite{PSOIwielly,nphotonbliokh2015spin}. Despite the considerable promise of these spin-orbit photonic effects, there remain several outstanding challenges to overcome both in terms of enhancing the relatively weak SOI effects, achieving control and tunability, and in terms of unique interpretation of the various exotic SOI effects that often manifest in a complex, interrelated way during an optical interaction \cite{nphotonbliokh2015spin,PSOIwielly,marruccinphoton}. Some inroads have already been made to enhance the SOI effects by tailoring a large spatial (or momentum) gradient of geometric phase (which is at the heart of SOI of light \cite{nphotonbliokh2015spin, hasmanscience}) in nanostructured metamaterials, metasurfaces \cite{capasoscience}, by using space-variant meta-polarization of light and through weak value amplification \cite{hosten2008observation}. Resonant enhancement and tunable resonance through hybridization of modes in plasmonic or in dielectric metamaterials are offering attractive means to achieve desirable control and tunability of the SOI effects \cite{capasoscience,hasmanscience,karimi2014generating,PSOIwielly}.\\
\\
Here, we report simultaneous observation of various exotic SOI effects in the interaction of highly non-paraxial light with a hybridized metamaterial, namely, waveguided plasmonic crystals (WPC). A custom-designed darkfield Fourier (momentum) domain polarization Mueller matrix experimental embodiment\cite{sciencemm,acsnano} enabled observation of the individual SOI effects in a decoupled manner despite their simultaneous occurrence in the WPC system. The waveguiding metallic photonic crystals comprising of a periodic metal nanostructures on top of a waveguiding dielectric layer, have been the subject of extensive recent research in the realm of nano-photonics. Their unique ability to couple and control both the electronic and the photonic resonances simultaneously through geometrical structuring has led to a plethora of interesting optical effects \cite{acsnano,christ2004optical,magneto}. The highly structured and strongly confined fields in such systems is also expected to lead to interesting SOI effects, which have not been investigated much\cite{vortexwpc}. We have thus explored such a possibility by precisely designing the geometric parameters so that the interference and hybridization of the waveguide resonance modes of the waveguiding dielectric layer and the surface plasmon resonance modes of the metal nanostructures in the WPC yields anisotropic (polarization-dependent) Fano resonance\cite{acsnano,pratunable,prageometric}. Due to the presence of the periodic metal nanostructures, the hybridized waveguide modes become leaky (quasiguided) and gets coupled to the radiation continuum. The interaction of tightly focused light with such leaky quasiguided modes of the anisotropic Fano resonant WPC system led to the exhibition of several non-trivial SOI phenomena in the far-field momentum domain. We have observed input spin (circular polarization)-depended opposite azimuthal trajectory of the quasiguided modes in the momentum domain, manifested as spin Hall effect of light. In the corresponding reciprocal effect, namely, the inverse spin Hall effect, the leaky quasiguided modes scattered at a particular azimuthal direction in the momentum space were observed to carry a specific spin polarization irrespective of the input state of polarization. Remarkably, the effect was observed even for incident unpolarized light  \cite{unpolshe,tsamunpol}. These and the other accompanying SOI effects are shown to arise due to the generation of space varying inhomogeneous polarization in the high numerical aperture (NA) imaging geometry and its subsequent interaction with the anisotropic WPC system. All the effects are separately manifested in different characteristic elements of the momentum domain Mueller matrix, enabling their interpretation in a single experimental embodiment. The anisotropic nature of the Fano resonance in the WPCs also enabled enhancement and control of the SOI effects. This is the first experimental demonstration of both forward and inverse spin Hall effect of light from the same system. Moreover, as opposed to previously observed spin-directional excitation of guided electromagnetic modes originating from three-dimensional nearfields \cite{soiwaveguide,soiwaveguide2}, here the leaky quasiguided modes enabled observation of these effects in the far-field through the transverse field components. \\
\\
\section*{Results \& Discussions}
\subsection*{Probing spin to orbital angular momentum conversion in plasmonic crystal using momentum domain Mueller matrix}
We have utilized a custom-designed polarization microscopic arrangement where one can determine the complete polarization response of a sample by recording the $4\times4$ Mueller matrix \cite{sciencemm,acsnano}(Figure 1a). This Mueller matrix imaging system integrated with a darkfield microscope employs broadband white light excitation and subsequent recording of thirty-six polarization resolved images of the sample-scattered light (at any selected wavelengths between $\lambda = 400 - 725 nm$) by sequential generating and analyzing six different linear and circular polarization states (see Materials $\&$ Methods)\cite{36mm}. Besides the real plane imaging and spectral Mueller matrix measurements, the Fourier plane Mueller matrix images (polarization-resolved transverse momenta distribution -$(k_x, k_y)$ can also be recorded at any selected wavelength using this experimental configuration. The darkfield polarization microscopic arrangement serves essential purposes in the emergence and probing of the SOI effects– (a) it enables the collection of the leakage light from the hybrid quasiguided modes of the WPCs bearing signature of the SOI due to scattering, which is otherwise relatively weak and swamped by the background light; (b) the highly non-paraxial imaging geometry enables efficient excitation of the quasiguided modes via large transverse momenta components $(k_x,k_y)$ of light, and (c) the high NA focusing also generates space varying inhomogeneous polarization at the sample plane, which mediates the SOI effects. \\
\\
The WPC samples consist of one-dimensional periodic gold (Au) gratings on top of an indium tin oxide (ITO) waveguiding layer, which is deposited on a glass substrate (Fig. 1b). The fabrication involved electron beam lithography and metal deposition by thermal evaporation technique (see Methods). The thickness of the waveguiding ITO layer is $190 nm$. The optimized dimension of the Au grating is (width $\approx 90 nm$, height $\approx 20  nm$, center to center distance $\approx 550  nm$).
In the WPC system, the transverse magnetic (TM) quasiguided modes in the dielectric waveguiding ITO layer and the surface plasmon resonances in the Au grating are simultaneously excited by the incident TM-$\boldsymbol{y}$ polarized light. Here, we have defined the direction of electric field for the TM-$\boldsymbol{y}$ polarization to be parallel to the direction of the grating vector $\boldsymbol{G}=2\pi/d$ (Fig. 1b)\cite{acsnano,christ2004optical,magneto}. Interference and hybridization of the discrete quasiguided modes and the broad surface plasmon mode lead to spectrally asymmetric Fano resonance\cite{christ2004optical,pratunable}. On the other hand, with TE- $\boldsymbol{x}$ polarization excitation (Electric field vector perpendicular to the grating vector), Fano resonance arises due to the interference of the TE quasiguided modes with the background photon continuum\cite{acsnano}. The geometrical parameters of the WPCs were judicially tailored (based on finite element method simulation\cite{fem}) so that the narrow TM and the TE quasiguided modes have strong spectral overlap and they lie within the surface plasmon resonance spectral band. Then, for either with TM or TE polarization excitation, the WPC exhibits ideal anisotropic Fano resonance (within the spectral range of $\lambda \approx 460 - 580 nm$) with strong amplitude anisotropy (described by linear diattenuation - $d_{wpc}$) and phase anisotropy (linear retardance -$\delta_{wpc}$) effects\cite{acsnano,prageometric}. The leakage radiation from the quasiguided modes \cite{beamingscience,PSOIwielly} is collected in our darkfield microscope arrangement to record both the polarization-resolved Fourier (momentum) domain Mueller matrix images and the spectral polarization response of the WPCs. The leakage radiation gets localized in the momentum domain as ring-like intensity distribution (Fig. 1c), which can be understood from the Fourier transformation of the fields corresponding to the quasiguided modes excited in the ITO waveguiding layer of the WPC \cite{oesimilar} (see supporting information S1). The radius of the diffraction rings in the far-field is decided by the wavevector of the quasiguided mode $|\boldsymbol{k_w}|$, and its width determined by the damping and losses of the modes. These diffraction rings in the far-field thus carry exclusive information on the near-field excitation and optical properties of the hybrid quasiguided modes\cite{beamingscience}. Note that in our experimental configuration, we could record only arc segments of the $+1$ and $-1$ orders of the diffraction rings as the spatial frequency in the momentum space $(k)$ is limited by the numerical aperture $(NA \approx 0.8)$ of the collection objective $(|\boldsymbol{k}|\leq k_0\times NA)$ (illustrated in Fig. 1c). The $4\times4$ Mueller matrix $(M)$ corresponding to these arc segments in the momentum domain were subsequently constructed  and the $1\times4$ Stokes vectors $\boldsymbol{S}$ of the sample-emerging light for a chosen input state are also recorded as a sub-set of the measurement process.\\ 
    \begin{figure}[!htp]
      \centering
      \includegraphics[scale=0.35]{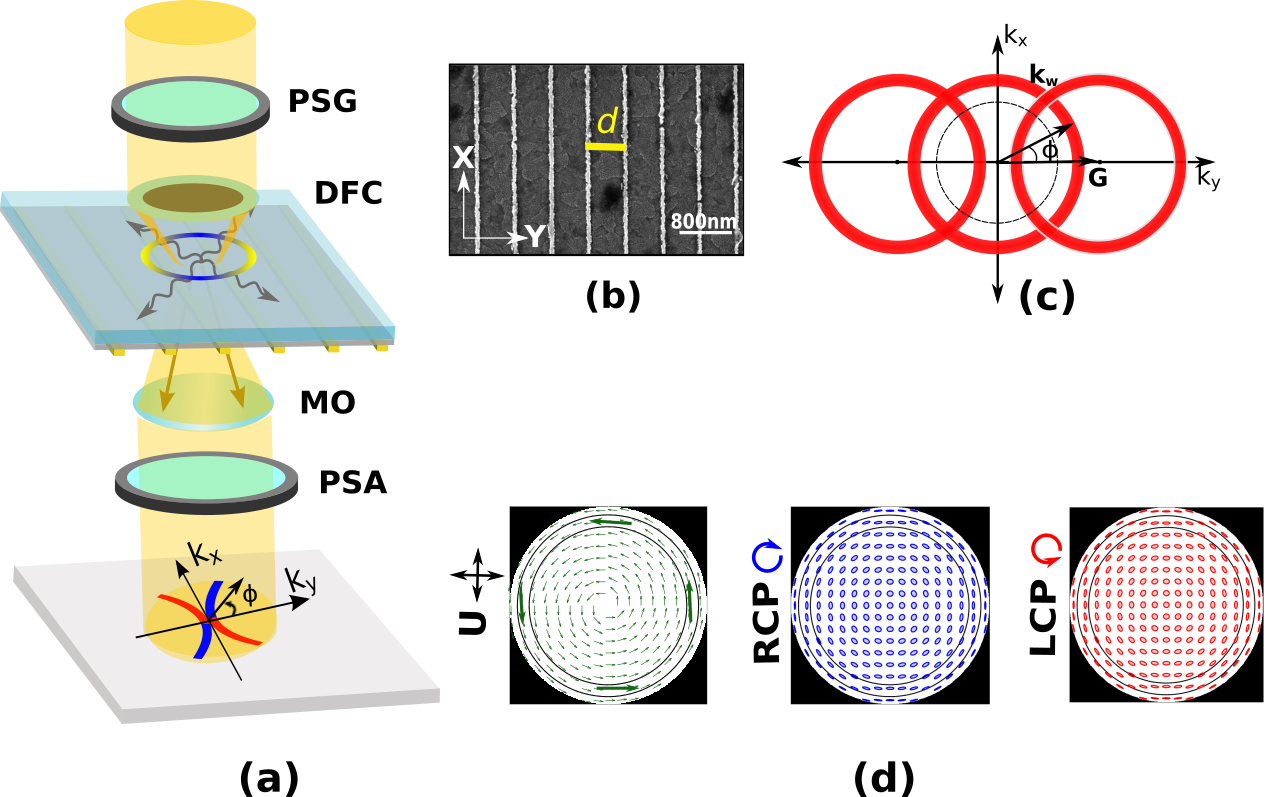}
      \caption{\textbf{Fig.1$\vert$ The signature of the leaky quasiguided photonic modes of the plasmonic crystals in the far-field momentum (Fourier) domain images.} (a) A schematic of the darkfield Fourier domain polarization Mueller matrix microscopy system. PSG and PSA: Polarization state generator and analyzer units, DFC: Darkfield condenser, MO: Microscope objective. (b) Typical SEM images of the Au grating waveguided plasmonic crystal. The direction of the TM-$\boldsymbol{y}$ and TE-$\boldsymbol{x}$ polarizations are shown. (c) The leakage radiation of the quasiguided modes manifest as ring-like intensity distribution in the momentum $(\boldsymbol{k})$ domain. The detected arc segments of the rings as permitted by the NA of the microscope objective are shown, and the radius of the rings corresponding to the wavevector $(\boldsymbol{k_w})$ of the quasiguided modes is illustrated. The azimuthal directional angle $(\pm\boldsymbol{\hat{\phi}})$ is defined with respect to the grating vesctor $\boldsymbol{G}$. (d) The theoretically simulated distribution of the space (azimuthal angle $\phi$ ) varying polarization of focused light at the sample site for incident unpolarized (U, green), RCP (blue), and LCP (red)  polarized light. The marked annular region corresponds to the numerical aperture of the darkfield condenser $(0.8-0.92)$ used in our experimental configuration.}
      \label{Fig. 1}
   \end{figure}\\
The spin to orbital AM coupling in non-paraxial optical fields (e.g., in tight focusing, imaging and scattering) can be described by the rotational transformation of the electric field polarization vector corresponding to the different wavevector $\boldsymbol{k}$ as\cite{nphotonbliokh2015spin}
\begin{equation}
     J=\begin{pmatrix}
     A+B \cos{2\phi} & B \sin{2\phi} & C \cos{\phi}\\
     B \sin{2\phi}&A-B \cos{2\phi}&C \sin{\phi}\\
     -C \cos{\phi}&-C \sin{\phi}&A+B\\
     \end{pmatrix}
    \end{equation}
Here, the SOI descriptor field coefficients $A, B, C$ encode information on the nature of the interaction (see Supporting information S2), and $\phi$ is the azimuthal angle in the case of cylindrically symmetric system \cite{nphotonbliokh2015spin}. Note that the SOI due to the propagation of paraxial light through a cylindrically symmetric inhomogeneous anisotropic medium can also be described using the  $2\times2$ sub-matrix corresponding to the transverse field components of Eq. (1). Here, we have a combination of both the effects as we have highly non-paraxial focused light interacting with the anisotropic Fano resonant WPC system, and the emerging transverse field components of the leaky quasiguided modes are imaged in the far-field. Tight focusing leads to the generation of azimuthally varying polarization of light at the WPC sample site for any input state of polarization (illustrated in Fig. 1d, see also Figure S1 of Supporting information S2). The propagation of such space varying polarization through the anisotropic WPC system having a fixed anisotropy axis (determined by the axis of the Au grating) leads to the generation of azimuthal geometric phase gradient, which mediates the spin to orbit AM conversions. The signature of the intrinsic SAM to intrinsic OAM (phase vortex) conversion mediated by the azimuthal geometric phase is evident in the $1^{st}$ $(S_1)$ and the $2^{nd}$  $(S_2)$ Stokes vector elements as characteristic “four-lobe” azimuthal intensity patterns in the $\boldsymbol{k}$- space arc segments of the quasiguided modes for input left (LCP) or right circularly (RCP) polarized light with helicity $\sigma = \pm1$ (Figure 2a). The results are shown for a wavelength $\lambda = 440 nm (\Delta \lambda \approx 40 nm)$, which is purposely selected away from the central region of the Fano resonance (shown subsequently). Clearly, the lobes arise due to the superposition of the original spin-polarized state $\sigma$ with the converted state $-\sigma$ having a geometric vortex phase factor of $2\sigma\phi$\cite{nphotonbliokh2015spin}. A similar signature of SAM to IOAM conversion is also evident in the 3rd Stokes vector element $(S_3)$ for input horizontally or vertically polarized light (Fig. 2b). \\
\\
The recorded $\boldsymbol{k}-$domain Mueller matrix $(M)$ (Fig. 2c) exhibits characteristics of a diattenuating retarder with azimuthally $(\phi)$ varying anisotropy axis, which encodes complete information on SOI in the anisotropic system (see Eq. 1 and Supporting information S2). While the elements$(M_{24/42},M_{34/43} )$ carry signature of SAM to IOAM conversion due to vortex linear retarder, $(M_{12/21},M_{13/31} )$ elements represent similar effects due to vortex linear diattenuator. The signature of the geometric phase evolution is evident in the $(M_{23/32})$ elements as an accumulation of $4\pi$ phase for a full $2\pi$ azimuthal rotation in the $\boldsymbol{k}$ space, confirming the phase singularity with a topological charge of 2. Interestingly, even the polarization-blind $M_{11}$ element also exhibits azimuthal lobe pattern $(\propto \sin^2{\phi})$ (Fig. 2c). This arises because even with unpolarized excitation, the high NA focusing generates an azimuthally varying linear polarization state at the WPC sample site (Fig. 1d). This accordingly excites azimuthally varying intensities of the TM and the TE leaky quasiguided modes. Surprisingly, the experimental $\boldsymbol{k}$ domain Mueller matrix exhibits non-zero magnitudes of the circular anisotropy descriptor $M_{14}$ and $M_{41}$ elements with characteristic azimuthal variations, which the conventional SOI matrix cannot explain (Eq. 1, Eq. E4 of S2 or similar formulations \cite{nphotonbliokh2015spin,waveoptics}). In what follows, we identify these as forward and inverse spin Hall effect of light and uncover its origin.\\ 
\begin{figure}[!htp]
    \centering
    \includegraphics[scale=0.2]{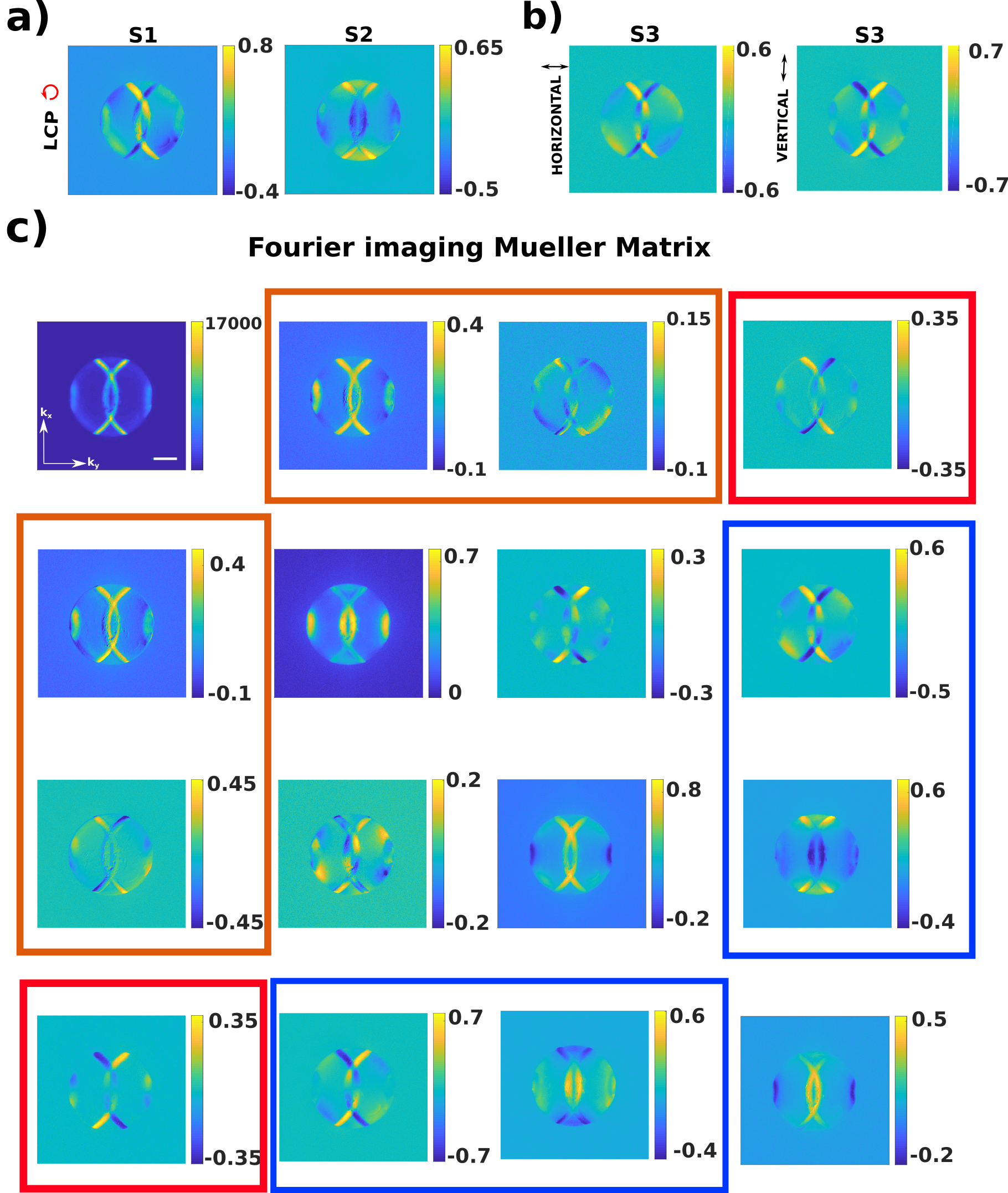}
    \caption{\textbf{Fig.2 $\vert$ Spin to orbital AM conversion of the quasiguided photonic modes of the WPC in the $\boldsymbol{k}$-space Stokes vectors and Mueller matrix elements.} Signature of the SAM to IOAM conversion through azimuthal geometric phase, manifesting as azimuthally varying intensity patterns in the $\boldsymbol{k}$-space arc segments, (a) in the Stokes vector elements $(S_1)$ and $(S_2)$ for input LCP state and (b) in the Stokes vector element $(S_3)$ for input horizontal and vertical polarization states. (c) The $\boldsymbol{k}$-domain Mueller matrix exhibiting characteristics of a diattenuating retarder with azimuthally $(\phi)$ varying anisotropy axis. The signature of SAM to IOAM conversion due to vortex linear diattenuator (elements marked by brown boxes) and vortex linear retarder (marked by blue boxes) are highlighted. Non-zero magnitudes of the circular anisotropy descriptor elements (red boxes) are apparent. The scale bar in the momentum space is $7\mu m^{-1}$.}
\label{Fig.2}
\end{figure}
\newpage
\subsection*{Forward and Inverse spin Hall effect of light and spin-directional coupling of the quasiguided modes in plasmonic crystal}
Forward spin Hall effect is manifested as input circular polarization (LCP/RCP) dependent “lighting-up” of the leaky quasiguided modes in opposite azimuthal directions $(\pm\boldsymbol{\hat{\phi}})$, generating spin-dependent intensity lobes in the $\boldsymbol{k}-$space arc segments (Figure 3a and 3b). In the inverse spin Hall effect, light scattered from the quasiguided modes along a given azimuthal  $\boldsymbol{k}$ direction carries a specific spin polarization, which is completely independent of the input state of polarization and is observed even for incident unpolarized light (Figure 3c, 3d and 3e). In terms of polarization effects, these represent azimuthally varying circular diattenuation (encoded in $M_{14}$) and circular polarizance effects $(M_{41})$, respectively. The origin of these effects can be unraveled by modeling the polarization evolution as sequential effects of - (i) generation of space varying inhomogeneous polarization due to focusing, which is described by a Mueller matrix of an azimuthal linear diattenuating retarder$ M_f(d_f, \delta_f, \phi)$; (ii) followed by its interaction with the anisotropic Fano resonant WPC, described by the Mueller matrix of a linear diattenuating retarder having a fixed orientation of the anisotropy axis as determined by the axis of the Au grating $M_{wpc}(d_{wpc}, \delta_{wpc}, \phi)$. The first one is a purely geometric effect, and the latter a resonant anisotropy effect. Using these transformations, the circular anisotropy descriptor $M_{14}$  and $M_{41}$  elements can be written as (see Supporting information S2)
\begin{equation}
\begin{split}
       M_{14}=-d_{wpc} \sqrt{1-d_f^2} sin (\delta_f) sin(2\phi)\\
      M_{41}=d_f \sqrt{1-d_{wpc}^2} sin (\delta_{wpc}) sin(2\phi) 
\end{split}
\end{equation}
Clearly, non-zero magnitude of the inverse spin Hall-descriptor $M_{41}$ element arises when $d_f\neq0$ and $\delta_{wpc}\neq0$. While linear diattenuation due to focusing $d_f\neq0$ is determined by the NA of the condenser \cite{waveoptics}, $\delta_{wpc}$ arises due to a phase difference between the TM and the TE hybrid quasiguided modes excited by orthogonal $\boldsymbol{(y-x)}$ linear polarizations \cite{acsnano}. Thus, for input unpolarized light, the quasiguided modes excited by  azimuthally varying linear polarization (Fig. 1d) will impart opposite circular (elliptical) polarizations to the leakage radiation at opposite azimuthal angles $(\pm\phi)$ with respect to the axis of the Au grating (as illustrated in Fig. 3c and results shown in 3d and 3e). This can be seen as a “geometric circular polarizer” comprising of an azimuthal linear polarizer followed by a waveplate, which yields the azimuthal circular polarizance effect of the quasiguided modes in the $\boldsymbol{k}$ space. The forward spin Hall-descriptor $M_{14}$, on the other hand, arises due to non-zero magnitudes of  $d_{wpc}$ and $\delta_f$. As previously discussed, differential excitation of the TM and the TE hybrid quasiguided modes by $\boldsymbol{y-x}$ linear polarizations lead to linear diattenuation effect $(d_{wpc}\neq0)$. The resulting azimuthal circular diattenuation effect can analogously be seen as a $\boldsymbol{k}$ space “geometric circular analyzer” that selectively leaks radiation from the quasiguided modes in opposite azimuthal directions $(\pm\boldsymbol{\hat{\phi}})$ for input LCP/RCP states. Note that the directionality of the leakage radiation is determined by the wavevector$\boldsymbol{(k_w)}$ of the quasiguided modes \cite{oesimilar} and the ring-like intensity patterns captured in the far-field Fourier domain carries information on this. Thus, the above-observed effects are  clearly far-field manifestation of the spin-directional excitation of the quasiguided modes in the WPC system. Indeed, inverse Fourier transformation of the intensity patterns corresponding to the recorded momentum domain Mueller matrix elements also supported this conclusion (see Supporting Figure S3).
We emphasize that the observation of the the forward and the inverse photonic spin Hall effects or the spin-directional coupling of the guided modes through the transverse field components in the far-filed momentum domain are exclusively enabled by the leaky quasiguided modes, which are accordingly amenable for interpretation using standard framework of Jones, Stokes-Mueller algebra. This is different from previous reports on similar effects which are typically ascribed to three-dimensional nearfield interference effects and transverse spin \cite{nphotonbliokh2015spin,PSOIwielly,soiwaveguide,ishenatcomm}. We further note that the above interpretations of the observed SOI effects were further comprehended by corresponding theoretical Mueller matrix analysis of the WPC system (see Supporting Figure S2).\\
\begin{figure}[!htp]
\centering
\includegraphics[scale=0.3]{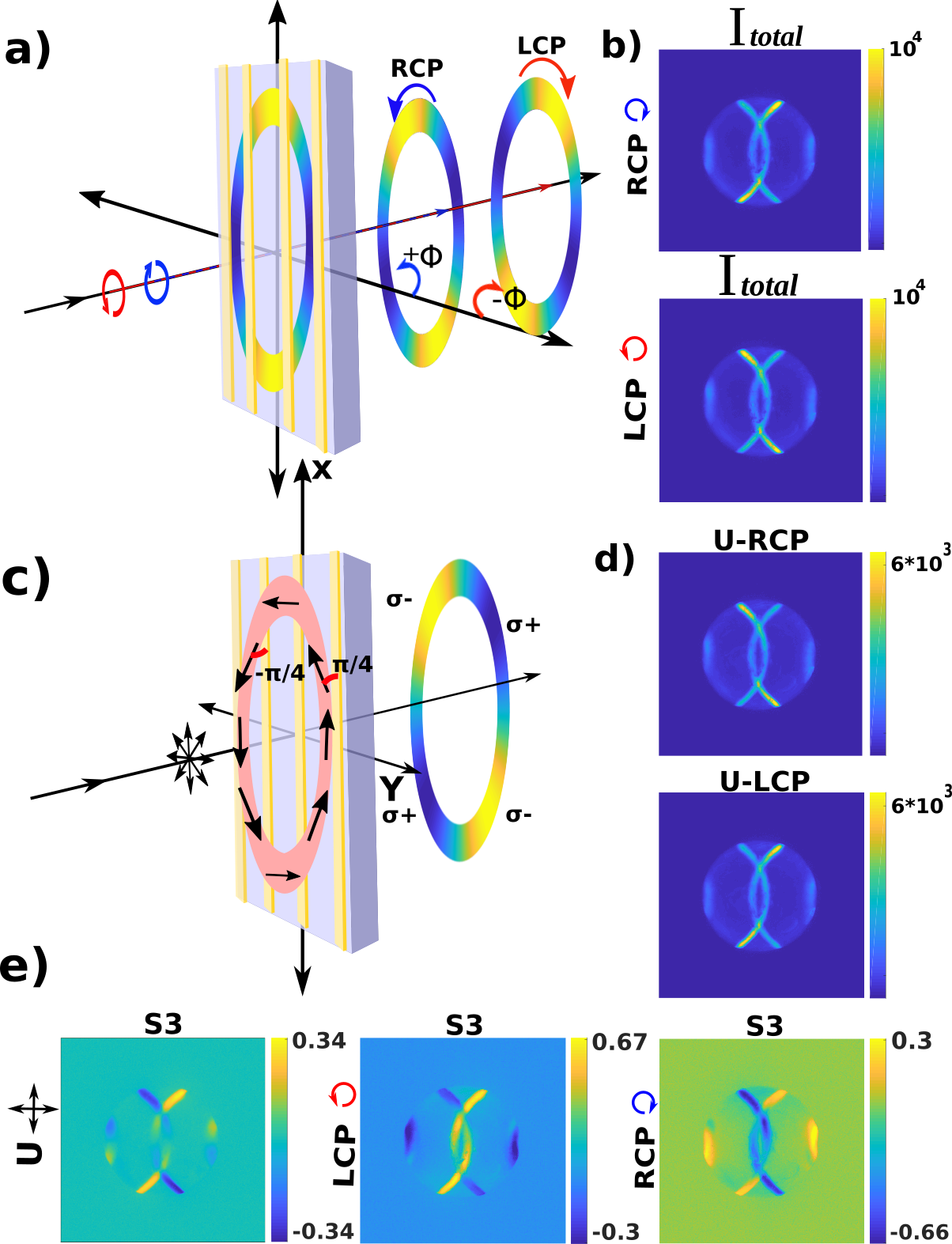}
\caption{\textbf{Fig.3 $\vert$ Forward and inverse spin Hall effect of the hybrid quasiguided modes of the plasmonic crystal.} Pictorial illustrations of the origin and appearance of the forward (a) and inverse (c) spin Hall effect of light scattered from the quasiguided modes. (b) The forward spin Hall effect is manifested as spin-dependent azimuthally $(\phi)$ varying intensity lobes on top of the total detected intensities for input LCP and RCP polarized light in the $\boldsymbol{k}$- space arc segments. (d) Inverse spin Hall effect is demonstrated as splitting of the opposite spin (circular polarizations) components of the scattered light at opposite azimuthal directions (for input unpolarized (U) light, marked as U-RCP and U-LCP). (e) The circular polarization descriptor Stokes vector element $(S_3)$ for input unpolarized, LCP, and RCP states. The circular polarization (spin) of the leaky quasiguided modes is observed to be selected by the azimuthal  $\boldsymbol{k}$ direction irrespective of the input state of polarization.}
\label{Fig. 3}
\end{figure}
\newpage
\subsection*{Fano resonance-enabled enhancement of the SOI effects}
It is evident from Eq. 2 that simultaneous occurrence of linear diattenuation and linear retardance $d_{wpc},\delta_{wpc}$ of the quasiguided photonic modes is responsible for simultaneous exhibition of both the SOI effects from the same WPC system and their strengths are also accordingly controlled by these parameters. Since these parameters are determined by the nearfield interference of the various resonant modes that leads to the anisotropic Fano resonance in the WPC system, these facilitate resonance-enabled enhancement of the SOI effects. This is demonstrated in Figure 4.  Prominent Fano resonance with characteristic asymmetric spectral line shape (in the spectral range $\lambda \approx 460 - 580 nm$, with a Fano dip around $\lambda = 490 - 500nm$) is observed for both TM $\boldsymbol{(y)}$ and TE $\boldsymbol{(x)}$  polarization excitations (Fig. 4a inset). While the Fano line shape corresponding to the TM polarization show slightly broader spectral features due to the excitation of the surface plasmon modes (which is not an ideal continuum), the corresponding spectra for TE polarization show much sharper asymmetric line shape. Never-the-less, the desired spectral overlap of Fano resonance between the TM and the TE polarizations leads to rapid increase of the Mueller matrix-derived $d_{wpc}$ and $\delta_{wpc}$ parameters (see Supporting information figure S4) around the Fano spectral asymmetry region (Fig. 4a), which is an exclusive signature of resonant enhancement of these anisotropy parameters \cite{acsnano,karimi2014generating,prageometric}. Fano resonance-enabled enhancement of both the photonic spin-hall effect and spin to orbital conversion (Fig. 4c) are accordingly observed in the characteristic azimuthal lobes in the $\boldsymbol{k}$- space arc segments of the quasiguided modes recorded at $\lambda=530 nm$, which falls within the central Fano spectral asymmetry region. The enhancement can be clearly seen in the recorded stokes vector $S_3$ for input unpolarized, LCP and RCP at $\lambda=530 nm$ as compared to that observed for $\lambda=440nm$ (Fig. 3e), which is away from the Fano spectral asymmetry region. While the magnitude of $S_3$ obtained with input unpolarized light demonstrate the enhancement of the inverse spin Hall effect, the enhancement of the SAM to IOAM conversion is confirmed from the measured Stokes vector element $S_3$ with input LCP and RCP states.\\
\begin{figure}[!htp]
\centering
\includegraphics[scale=0.4]{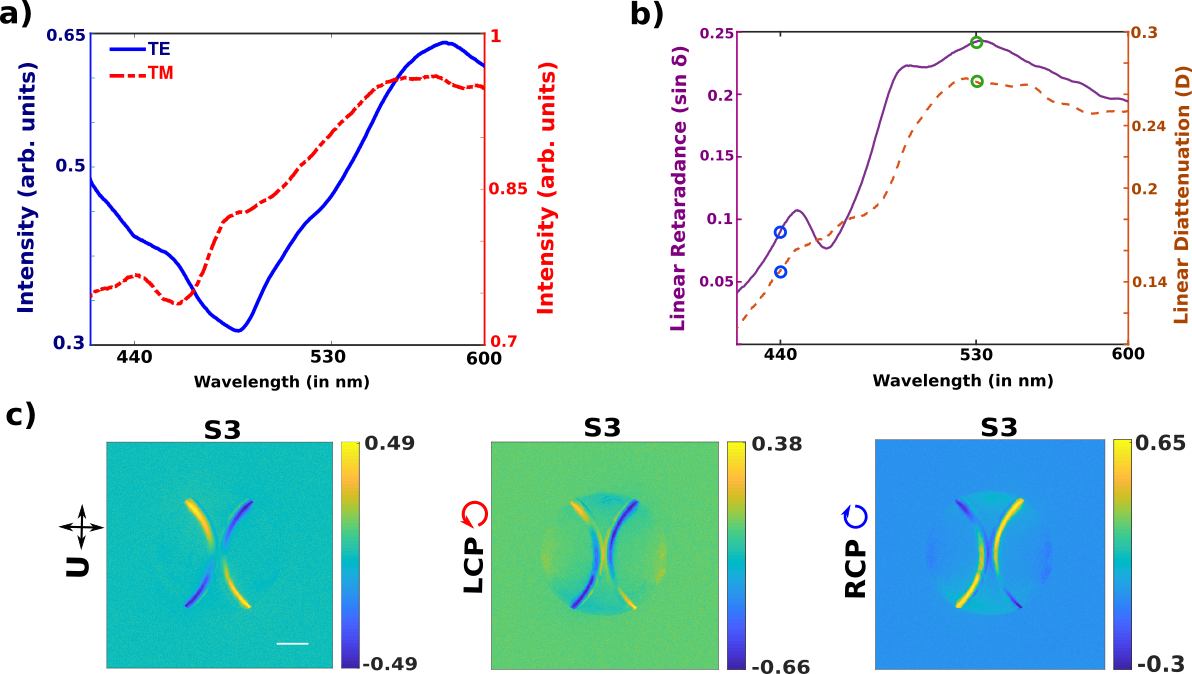}
\caption{\textbf{Fig.4 $\vert$ Fano resonance-enabled enhancement of the SOI effects.} (a)The scattering spectra from the leaky quasiguided photonic modes of the Au grating WPC for TM-$\boldsymbol{y}$ (red dotted line) and TE-$\boldsymbol{x}$ (solid blue line) polarization excitation exhibit prominent Fano spectral asymmetry with considerable overlap $( \lambda \approx 460 - 580 nm)$. (b) The anisotropic nature of Fano resonance and resonant enhancement of anisotropy is evident from the rapid variation of the Mueller matrix-derived linear retardance-(solid violet line) and linear diattenuation- (brown dotted line) parameters around the Fano spectral asymmetry region $( \lambda \approx 460 - 580 nm)$. (c) The Stokes vector element $S_3$ of the $\boldsymbol{k}$-space arc segments of the leaky quasiguided mode recorded for input unpolarized (U), LCP and RCP light at $\lambda=530 nm$ demonstrates resonant enhancement of froward and inverse spin Hall effect of light. Depicted momentum domain Scale bar is $5\mu m^{-1}$.}
\label{fig4}
\end{figure}
\newpage
\section*{Conclusion}
In conclusion, we have observed spectacular manifestation of both the forward and the inverse spin Hall effect of light through the excitation of the quasiguided photonic modes in a plasmonic photonic crystal metamaterial. The leaky quasiguided modes provided a unique means to study the input spin-dependent directional scattering phenomena and its reciprocal effect of direction (wavevector)-controlled spin selection of light in the far-field. These phenomena were manifested in the momentum domain Mueller matrix elements, enabling their interpretation as geometrical circular anisotropy effects. Nearfield interference of the modes also offered Fano resonance-enabled enhancement of the SOI effects. Simultaneous exhibition of spin-directional coupling of the quasiguided modes and other accompanying SOI effects in the far-filed, observation of the spin-orbit coupling phenomena even for completely unpolarized light, characteristic manifestations of all the effects in the momentum domain Mueller matrix, the ability to regulate and control the SOI effects in geometrically tailored plasmonic crystals, altogether open up exciting avenues of spin-orbit photonic research. It is envisaged that the possibility of simultaneously realizing spin-controlled directional guiding of waves and direction-dependent control of angular momentum of light in the same system of plasmonic photonic crystal will enhance the scope of spin-orbit photonic applications in directional switching, polarization sorting, processing of polarization-encoded information and in developing novel nanophotonic sensors. \\
\\
\section*{Materials and Methods}
\subsection*{Darkfield Mueller matrix imaging and spectroscopy system}
Both the spectral Mueller matrices and Fourier domain Mueller matrix images from the plasmonic nano-structure are recorded using a home built darkfield Mueller matrix system\cite{sciencemm,acsnano}. Essentially the darkfield microscope consists of a conventional inverted microscope (Olympus, IX71) with a darkfield condenser (DFC) (olympus U-DCD, numerical aperture (NA) $0.8-0.92$), which facilitates efficient collection of exclusively the scattered light form nano-structures. The Mueller matrix measurement strategy is integrated with this microscope by sequentially generating and analysing six different linear and circular polarization states. For this purpose, the collimated white light from the microscope's inbuilt illumination source (halogen lamp, JC 12V 100W) is passed through a polarization state generator unit which comprises of a rotatable (i) linear polarizer and (ii) achromatic quarter wave plate (QWP, Thorlabs AQWP05M-600 ). The light is then focused to an annular shape at the sample site by the DFC and the scattered light from the sample is collected by the microscope objective (Olympus,MPlanFL, NA $= 0.8$). Polarization of the scattered light is subsequently analyzed by the polarization state analyzer unit that consists of a rotatable achromatic QWP and a linear polarizer. All the optical polarizing elements are mounted on computer controlled motorized rotational mounts (PRM1/M-Z7E, Thorlabs, USA) for precision control. For recording the spectral Mueller matrices of the WPCs, the scattered light is relayed to a spectrometer (HR 4000, Ocean Optics) for spectrally resolved signal detection (resolution $.2nm$). The  Mueller matrix images at selected wavelengths were recorded by using an EM CCD camera (Andor, IXON 3). The Fourier domain images were recorded using a Fourier transforming lens (f $= 75 mm$). The selection of the spectral regions of interest were achieved using appropriate band-pass filters (Blue:- BP $400-440 nm$), (Green:- BP $510-550 nm$).  Both the spectral and the Fourier imaging Mueller matrix were constructed using 36 polarization-resolved projective measurements (given in the table below)\cite{36mm}.
\begin{table}[!htp]
\centering
\resizebox{\textwidth}{!}{
\begin{tabular}{|c|c|c|c|} 
\hline
HH+HV+VH+VV & HH+HV$-$VH$-$VV & PH+PV$-$MH$-$MV & RH+RV$-$LH$-$LV\\
\hline
HH$-$HV+VH$-$VV & HH$-$HV$-$VH+VV & PH$-$PV$-$MH+MV & RH$-$RV$-$LH+LV\\
\hline
HP+VP$-$HM$-$VM & HP$-$VP$-$HM+VM & PP$-$PM$-$MP+MM & RP$-$RM$-$LP+LM\\
\hline
HR+VR$-$HL$-$VL & HR$-$VR$-$HL+VL & PR$-$PR$-$MR+ML & RR$-$RL$-$LR+LL \\
\hline
\end{tabular}}
\caption{ \textbf{Table 1 $\vert$} Scheme for construction of  Mueller matrix using 36 polarization-resolved projective measurements. Here, the first letter represents the input polarization state, the second letter stands for the analyzer or the projected polarization state. The states are defined as $I_H(horizontal)$, $I_V(vertical)$, $I_P (+45\deg)$, $I_M(-45\deg)$, $I_L$ left circular polarized $(LCP)$, $I_R(RCP)$  }
\label{table:Mueller}
\end{table}
\subsection*{Fabrication of the waveguided plasmonic crystals} The hybrid plasmonic crystal consists of $1D$ periodic gold gratings on top of an ITO waveguide layer with thickness $\approx 190nm$ which is deposited on a glass substrate. The gratings  are fabricated with periodicity $\approx 550nm$, width $\approx 90nm$ and thickness $\approx 20nm$. The nanostructure is fabricated using the electron beam lithography (Zeiss SIGMA field emission microscope) and the gold metal is deposited using thermal evaporation technique.The thickness of the ITO waveguide layer and geometric parameters of the grating are chosen in away that enables simultaneous excitation of the plasmonic and hybrid waveguide modes in the desired spectral range $(\lambda \approx 420 - 600 nm)$.\\
\newpage
\section*{Supporting information}
\subsection*{S1: Manifestation of the leaky quasiguided modes of the plasmonic crystals in the far-field momentum domain}
The nanostructured metamaterial investigated in this study is a metallic photonic crystal that provides simultaneous excitation of both the plasmonic and photonic resonances. With input transverse magnetic (TM) polarization, where the electric field is perpendicular to the direction of the Au grating axis, excitation of the plasmon resonances of the Au grating and the TM waveguide modes in the ITO waveguiding layer leads to the formation of hybridized waveguide plasmon polaritons. For TE excitation, an asymmetric Fano-type line shape is observed in the scattering spectra, which emerges due to the coupling of the waveguide modes with the incoming photon continuum\cite{acsnano,christ2004optical}. In both scenarios, the guided modes become leaky due to the periodic metallic grating. This leakage radiation from the quasiguided modes  collected by the microscope objective is used to obtain both the Fourier domain images and the scattering spectra of the waveguided plasmonic crystal (WPC). Here the high numerical aperture (NA) of the dark-field condenser (DFC) provides additional transverse momenta $(k_x,k_y)$ to the input non-paraxial beam to efficiently excite the waveguide modes\cite{christ2004optical,pratunable}. The electric field distribution of the hybrid quasiguided modes will act as the source for the far-field (momentum domain) response of the WPC system through the leakage radiation. The optical disturbance at the Au grating WPC interface associated with the electric field of the leakage radiation can be written as\cite{oesimilar} 
    \begin{equation}
        U_{\phi, p}(x,y,z=0)= C_{\phi}[1+sin (2\pi x/d)]e^{i\Vec{k}_{w}.\Vec{r}}
    \end{equation}\\
where $C_{\phi}$ is an amplitude constant, $\boldsymbol{k_w}$ is the wavevector of the waveguide modes, $\phi$ is the azimuthal angle and $d$ is the periodicity of the Au grating. Consequently, the far-field response can be obtained through Fourier transformation of the above amplitude distribution as 
     \begin{equation}
    \begin{split}
        U_{k_x,k_y} \propto &\iint C_{\phi}[1+sin (2\pi x/d)]e^{i(k_{w}cos\phi +k_{w}sin\phi)}e^{-i(k_{x}.x+k_{y}.y)} \,dx\,dy
    \end{split}
    \end{equation}
    \begin{equation}
    \begin{split}
        U_{k_x,k_y} \propto &\delta (k_x-k_w cos\phi, k_y -k_w sin \phi) \\ &\delta (k_x-(k_w cos\phi+G), k_y -k_w sin \phi) \delta (k_x-(k_w cos\phi-G), k_y - k_w sin \phi)
    \end{split}
    \end{equation} 
\\
It is evident that in the Fourier plane the intensity distribution will appear as circular diffraction rings having infinitesimal width (described by the $2D$ delta function) and with different orders separated by the grating vector $\boldsymbol{G}=2\pi/d$. Here, the wavevector of the quasiguided mode $|\boldsymbol{k_w}|$ will decide the radius of the rings. In  experimental observations, rings with finite thickness are observed due to the inherent damping and losses of the hybrid quasiguided modes, which is not considered for simplicity in Eq.2. A pictorial illustration of the observed intensity distribution provided in Figure. 1(c) of the main text describes the formation of the zeroth and the first $(\pm 1)$ order diffracted rings. In our experimental configuration, with finite NA of the microscope objective ($\text{NA}=0.8$), only some portion of the $+1$ and $-1$ diffracted rings are collected as $|\boldsymbol{k_w}|>|k_0 \cdot \text{NA}|$ \& $|\boldsymbol{k_w}-\boldsymbol{G}|<|k_0 \cdot \text{NA}|$ (Fig. 1c of the main text). This leads to the appearance of the arc segments in the momentum domain intensity images. Unlike the azimuthally invariant excitation of the surface waves\cite{oesimilar}, we have observed lobed intensity distribution in the diffraction rings due to the directional excitation of the quasiguided modes in the WPC, as described in the main text. The signature of the various SOI effects in the far-field momentum domain are subsequently captured by recording the polarization-resolved intensity images of these two arcs of the first-order diffraction rings. It may be pertinent to note that this type of measurements using leaky quasiguided modes offers the possibility of capturing information with sub-diffraction (sub-wavelength) resolution\cite{oesimilar}, which might be of general importance towards studying SOI effects in nanostructured metameterials.\\
\subsection*{S2: Mueller matrix modelling of SOI due to tight focusing and subsequent excitation of the quasiguided photonic modes in the WPC}
It is well known that tight focusing using high NA lens rotates the incoming wave-vectors in the meridional plane, generating a conical $\boldsymbol{k}$ distribution, which in turn leads to the rotation of the associated local polarization vectors to satisfy the transversality condition\cite{nphotonbliokh2015spin}. This leads to the evolution of the geometric phase and its gradient which mediates the spin-orbit interactions (SOI) of light. The evolution of the polarization vectors can be calculated using the $3d$ polarization transformation matrix (Eq.1 of the main text) using the Debye-Wolf theory\cite{deby-wolf}. It is pertinent to note that the polarization distribution of the real space electric field near the focal plane can be obtained using the so-called Debye-Wolf integral, which is obtained by integrating over all the available wave-vectors provided by the NA of the focusing lens. In the  Debye-Wolf theory and in other related approaches for describing SOI of light in tight focusing, the corresponding polarization transformation matrix is expressed in terms of the well known Debye-Wolf diffraction integrals $I_0, I_1$ and $I_2$ \cite{deby-wolf}. Note that the polarization transformation in the $\boldsymbol{k}$- space can be modelled using the $\boldsymbol{k}$-resolved rotation matrix as given in Eq.1 of the main text, where integration over different $\boldsymbol{k}$ is not needed. Also, as we have recorded the scattered far-field intensity pattern in the imaging geometry, it is sufficient to consider only the transverse field components. Thus, the corresponding matrix will be a conventional $2\times2$ Jones matrix by considering only the first two rows and columns of the matrix given in Eq.1, where the coefficients $A, B, C$ are given by $A=\frac{(1+cos(\theta))}{2}$, $B=\frac{(1-cos(\theta))}{2}$, $C=sin(\theta)$ where $\theta$ is the focusing angle \cite{waveoptics}. We would also like to note that irrespective of the $\boldsymbol{k}$ integrated (real space) or $\boldsymbol{k}$ resolved (momentum space) measurements, the obtained Mueller matrix exhibits the characteristics of a linear diattenuator with azimuthally varying orientation axis, which can be written as
    \begin{equation}
    M_f = \begin{pmatrix}
    1 & d_f \cos{2\phi} & d_f \sin{2\phi} & 0\\
    d_f \cos{2\phi} & \cos^2{2\phi}+x_f \sin^2{2\phi} & (1-x_f) \cos{2\phi} \sin{2\phi} & 0\\ d_f \sin{2\phi} & (1-x_f) \cos{2\phi} \sin{2\phi} & \sin^2{2\phi}+x_f \cos^2{2\phi} & 0 \\ 0 & 0 & 0 & x_f\\
    \end{pmatrix} 
   \end{equation}
    Here, $d_f$ is the linear diattenuation due to the focusing and $x_f = \sqrt{1-d_f^2}$.
   \begin{figure}[!h]
     \centering
     \includegraphics[scale=0.5]{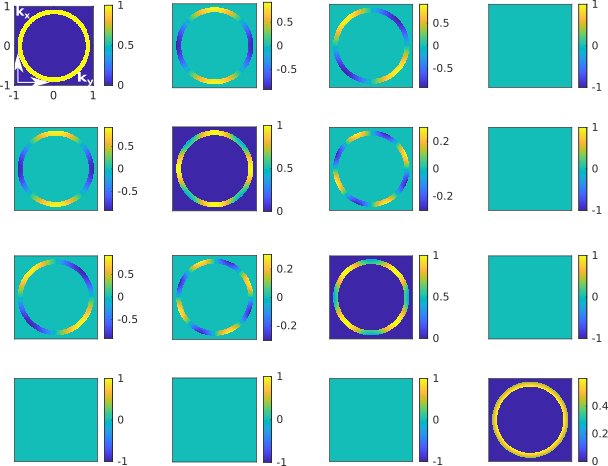}
    \caption{\textbf{Fig.S1 $\vert$ Mueller matrix characterization of spin orbit interaction due to tight focusing.}  Theoretically generated Mueller matrix corresponding to high NA focusing. The matrix represented in the focal (Fourier) plane shows characteristics of a linear diattenuator having azimuthally varying orientation axis. The annular region corresponds to the numerical aperture $(0.8-0.92)$ of the dark-field condenser used in our experiment.}
     \label{Fig.S2}
  \end{figure}
\\
As previously discussed, the Mueller matrix of the WPC has the form of a linear diattenuating retarder with a fixed orientation axis determined by the orientation axis of the Au grating of the WPC. In the experimental system, the WPC is placed such a way that the  grating axis is parallel to the horizontal axis of the laboratory linear polarizer in the polarization state generator unit. In this configuration, the Mueller matrix of the WPC can be written as
   \begin{equation}
     M_g =\begin{pmatrix}
          1 & d_{wpc} & 0 & 0\\
          d_{wpc} & 1 & 0 & -sin(\delta_{wpc})\\
          0 & 0 & \sqrt{(1-d_{wpc}^2)} cos(\delta_{wpc}) & 0\\
          0 & sin(\delta_{wpc}) & 0 & cos(\delta_{wpc})\\
          \end{pmatrix}
   \end{equation}
We have modeled the polarization transformation corresponding to the SOI effect in the far-field as sequential product of the Mueller matrices corresponding to the polarization transformation due to focusing and due to the anisotropic Fano resonant WPC system. It is pertinent to note that although the Mueller matrix corresponding to pure focusing transformation is ideally an azimuthal linear diattenuator (Fig. S1), nonzero linear retardance effect may also arise in practical scenario due to refraction or reflection effects at interfaces (which was the case in our experiments). Thus, in our case, the resultant Mueller matrix describing the far-field SOI effect is described as the product of two matrices, both of which are linear diattenuating retarders, with one having an azimuthally varying orientation axis and the other having a fixed orientation of the anisotropy axis. 
   \begin{equation}
       M=  M_f(d_f, \delta_f, \phi)\times M_{wpc}(d_{wpc}, \delta_{wpc}, 0)
   \end{equation}\\
   The theoretically generated Mueller matrix corresponding to our experimental measurements is presented in Figure S2.
    \begin{figure}[!h]
      \centering
      \includegraphics[scale=0.5]{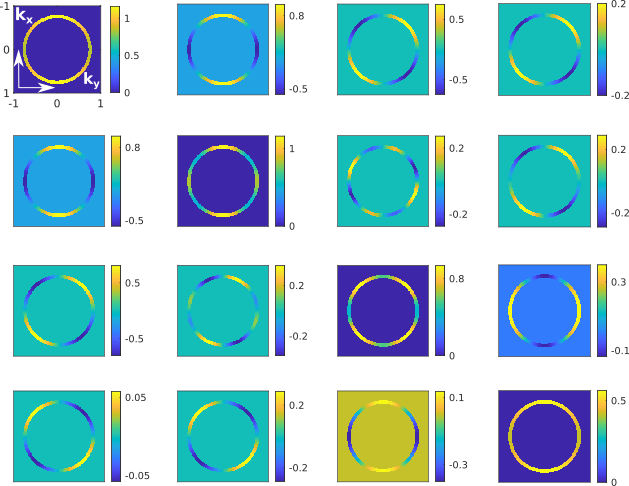}
     \caption{\textbf{Fig S2 $\vert$ Theoretical Mueller matrix of the waveguided plasmonic crystal.} The theoretically generated Mueller matrix of the anisotropic WPC upon excitation with tightly focused light in our experimental configuration. The results are for a representative wavelength $\lambda \approx 530nm$. The modelling parameters are taken comparable to our experiment. NA of the DFC is $0.8-0.92$ and the linear diattenuaton and linear retardance parameters used in the simulation are the experimental values of the WPC at $\lambda \approx 530nm$. Besides the usual manifestation of spin to orbital angular momentum conversion in the characteristic Mueller matrix elements, the non-uniform azimuthal intensity distribution in $(M_{11})$ is observed owing to the directional excitation of the hybridized waveguide-plasmon polariton modes. In conformity with the experimental results, both the forward and the inverse spin-hall effects are observed in the $M_{14}$ and $M_{41}$ elements respectively.}
      \label{Fig.S3}
    \end{figure}
\newpage
\subsection*{S3: Evidence of spin-controlled directional excitation of the leaky quasiguided modes in the plasmonic crystal}
As shown in the main text, the leaky quasiguided  modes of the WPC manifested in the momentum (Fourier) domain as circular ring-like intensity distributions and the different SOI effects in the far-field revealed their signature as polarization-dependent azimuthal intensity lobes on top of it. Thus, the leaky quasiguided modes carried the SOI information on the forward and the inverse spin Hall effects of the guided modes of the WPC into the far-field momentum domain. Although, quantitative understanding on the exact nature of the spin-directional excitation of the quasiguided modes will necessitate measurements of three dimensional nearfileds, some qualitative information on this may also be obtained by performing inverse Fourier transformation of the far-field intensity distributions of selected momentum domain Mueller matrix elements. Indeed,  all the Fourier-processed Mueller matrix elements showed signature of the waveguide mode patterns (not shown). The signatures of circular polarization (spin)-dependent directional excitation of the quasiguided modes and it's reciprocal effect (wavevector-dependent selection of spin polarization of the modes) are also evident in the corresponding Fourier-processed intensity data, as shown in Fig. S4 (a) and (b), respectively. \\
    \begin{figure}[!h]
      \centering
      \includegraphics[scale=0.5]{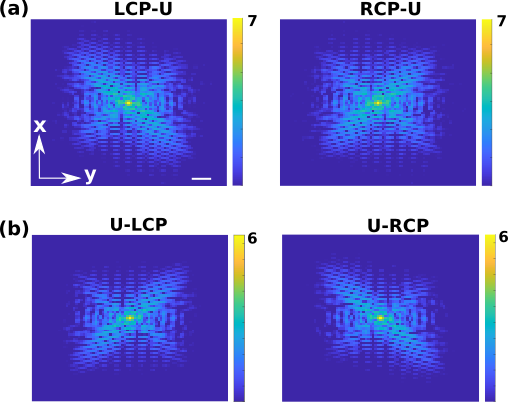}
     \caption{\textbf{Fig.S3 $\vert$ Spin-directional coupling of the leaky quasiguided photonic modes of the plasmonic crystal.}  Inverse Fourier transformation of the intensity patterns of selected momentum domain Mueller matrix elements  of the leaky quasiguided modes provides evidence for the spin-directional excitation of the quasiguided modes. The radially decaying quasiguided mode patterns are visible in all the images. (a) The spin-dependent directional excitation of the hybrid quasiguided modes is observed in the Fourier processed total intensities for LCP and RCP input light, respectively (denoted as LCP-U and RCP-U), (b) The reciprocal effect of direction-dependent spin selection is also observed in the LCP and RCP components for input unpolarized light (denoted as U-LCP and U-RCP). The shown scale bar is $8\mu m$.}
      \label{Fig.S4}
    \end{figure}\\
\subsection*{S4: Signature of anisotropic Fano resonance in the spectral Mueller matrix of the  plasmonic crystal}
The geometrical parameters of the WPC was tailored so that it exhibits prominent anisotropic Fano resonance in the wavelength region $\lambda \approx 460 - 580 nm$. The anisotropic Fano resonance-enabled enhancement of both forward and inverse spin Hall effect of the quasiguided modes of such geometrically tailored WPC system are demonstrated in Figure 4 of the main text. The enhancement of the SOI effects are mediated by the resonant-enhancement of the Fano anisotropy parameters namely, linear diattenuation $(d_{wpc})$ and linear retardance $(\delta_{wpc})$ parameters. Here, in Figure S4, the corresponding full spectral Mueller matrix from the anisotropic WPC sample is shown. The recorded Mueller matrices were subjected to polar decomposition analysis and the linear diattenuation and the linear retardance parameters were subsequently derived using standard algebraic manipulations\cite{acsnano,polarlu1996interpretation}.   
    \begin{figure}[!h]
  \centering
  \includegraphics[scale=0.5]{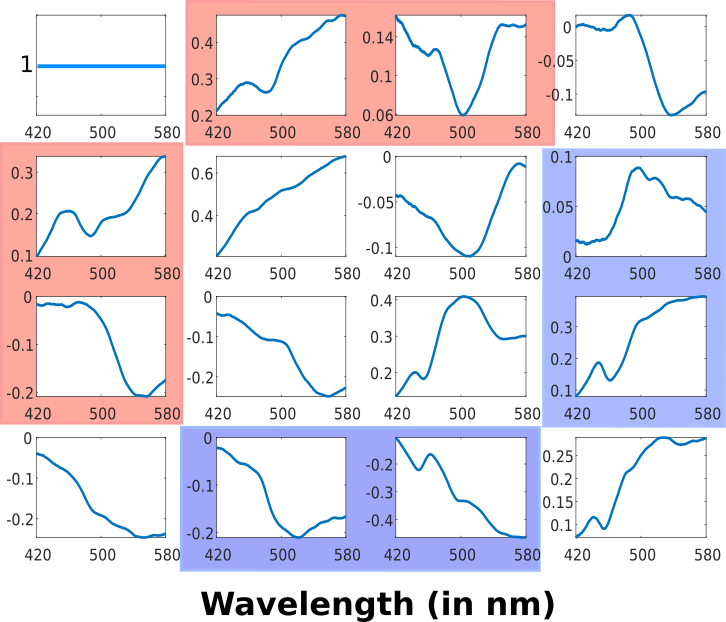}
 \caption{\textbf{Fig.S4 $\vert$ Signature of anisotropic Fano resonance in the spectral Mueller matrix elements.}  The experimental scattering Mueller matrices of the waveguided plasmonic crystal are shown as a function of wavelength ($\lambda= 420 - 600 nm$). The elements are normalized with respect to the $M_{11}$ element. The obtained spectral Mueller matrix exhibits characteristics of a spectrally varying diattenuating retarder. The linear diattenuation and linear retardance descriptor elements are highlighted in brown and blue boxes, respectively. Rapid variations of these elements around the Fano spectral asymmetry region is a characteristic signature of anisotropic Fano resonance.}
      \label{Fig.S1}
\end{figure}
\newpage
\bibliographystyle{unsrt}
\bibliography{main}
\end{document}